\documentstyle[12pt, aasms4]{article}

\slugcomment{To be submitted to ApJ}

\lefthead{Nishiuchi et al.}
\righthead{{\it ASCA} observations of GRO J1744--28} 

\begin{document}
\newcommand {\gro}{{GRO~J1744--28}}
\newcommand {\etal}{{\it et~al.}}
\newcommand {\ergcms} {erg cm$^{-2}$ s$^{-1}$}
\newcommand {\Msun} {{\it M$_{\odot}$}}
\newcommand {\ergs} {erg s$^{-1}$}

\title{ASCA Observations of GRO J1744--28}

\author{M. Nishiuchi\altaffilmark{1}, K. Koyama\altaffilmark{2} and 
Y. Maeda\altaffilmark{3}}
\affil{Department of Physics, Graduate School of Science, Kyoto University, 
Sakyo-ku, Kyoto, 606-8502, Japan} 
\altaffiltext{1}{E-mail address: mamiko@cr.scphys.kyoto-u.ac.jp}
\altaffiltext{2}{CREST, Japan Science and Technology Corporation (JST), 
4-1-8 Honmachi, Kawaguchi, Saitama, 332, Japan}
\altaffiltext{3}{Department of Astronomy and Astrophysics, 525 Davey 
Laboratory, Pennsylvania State University, University Park, PA 16802, U.S.A.}

\author{K. Asai,  T. Dotani,   H. Inoue,  K. Mitsuda,  F. Nagase,  Y. Ueda}
\affil{ Institute of Space and Astronautical Science, Yoshinodai, 3--1--1,
 Sagamihara, Kanagawa 229-8510, Japan}

\and

\author{C. Kouveliotou}
\affil{Universities Space Research Association, at ES-84, NASA/MSFC, 
Huntsville, AL 35812, U.S.A.}

\begin{abstract}

We report the ASCA results of the bursting X-ray pulsar \gro,
which was observed in February 1996 and March 1997.
The source flux in the 2--10 keV band was $2.0\times10^{-8}$ 
erg/sec/cm$^2$ in 1996 and
$5.0\times10^{-9}$ erg/sec/cm$^2$ in 1997.
We detected 12 and 17 Type II bursts during the two observations
with mean bursting intervals of about 27 min and 37 min.
Each burst is followed by an intensity dip with the depleted flux 
depending on the burst fluence. 
The energy spectra are approximated by an absorbed power law
with additional structure around 6--7 keV\@.
Constant absorption column, $(5-6)\times10^{22}$ cm$^{-2}$, 
independent of the observation dates and emission phases (persistent, burst
and dip) is interpreted as an interstellar absorption.
The source may be actually located near the Galactic center, at
a distance of 8.5 kpc.
The structure in the energy spectrum at 6--7 keV is most probably
due to iron and maybe reproduced by a disk line model
with additional broadening mechanism.

\end{abstract}


\section{Introduction}

\gro, an X-ray transient bursting pulsar,
was discovered near the Galactic center on 1995 Dec.\ 2
with the Burst And Transient Source Experiment (BATSE)  
on board the Compton Gamma-Ray Observatory (CGRO)
(Fishman \etal\ 1995;  Kouveliotou \etal\ 1996). 
Using the BATSE data, Finger \etal\ (1996) concluded that 
\gro\  is a low mass X-ray binary system including a neutron 
star with pulse frequency of 2.14~Hz and a binary 
orbital period of 11.83 days; the mass function of the
system is 1.36 $\times$ 10$^{-4}$ M$_{\odot}$.
 Finger \etal\ (1996) estimated
the magnetic field to be $<6\times10^{11}$ G\@. 
Later, Cui \etal\ (1997) analyzed observations with
the Proportional Counter Array (PCA) on Rossi X-ray
Timing Explorer (RXTE)
and estimated the magnetic field strength to be
$\sim~2.4\times10^{11}$~G,
assuming that the system was in the propeller state.
Zhang \etal\ (1996) discovered Quasi-Periodic
Oscillations (QPO) of the X-ray intensity at 20~Hz, 40~Hz and 
60~Hz in the PCA data; 
QPO features at 20~Hz and 60~Hz were 
weak, while the 40~Hz QPO was quite prominent.

Major outbursts have been detected twice so far from \gro.
The first outburst started in December 1995, which lead to
the discovery of the source.
The burst activity was terminated in early May 1996
(Kouveliotou \etal\ 1996b).
The second outburst began about a year later in December 1996
(Kouveliotou \etal\ 1997) and ended in April 1997.
Both outbursts decayed with a time scale of a few tens of days. 
In addition to these major outbursts \gro\ seems to show
small outbursts with shorter
durations (Jahoda \etal\ 1996), detectable only below 50 keV.

Since the discovery of \gro, several thousands of bursts have been observed.
Following some bursts, the flux decreases below the persistent level for 
a few seconds to a few minutes (hereafter referred to as a dip), 
depending on the burst fluence (Swank \etal\ 1996). 
Kouveliotou \etal\ (1996) attributed these bursts to accretion
instabilities; similarly, Lewin \etal\ (1996) remarked that
these bursts were probably due to
spasmodic release of gravitational potential energy (Type II) rather 
than thermonuclear burning (Type I), comparing the light curves 
measured by the RXTE with those of the Rapid Buster.
Bildsten and Brown (1997), however, suggested that the bursts from
\gro\ during its low persistent flux level could be due to the 
thermonuclear burning because the neutron star in \gro\  has 
only a weak magnetic field.

A shift in the arrival time of the pulse during the bursts was 
observed by the Oriented Scintillation Spectrometer Experiment 
(OSSE; Blanco \etal\ 1996, Strickman \etal\ 1996) 
and BATSE (Koshut \etal\ 1998) on board the CGRO and PCA on
board RXTE (Stark \etal\ 1996).
Strickman \etal\ (1996) reported  on the phase shift observed during
the bursts, while Stark \etal\ (1996) reported on the pulse arrival 
time lag which
persisted longer than the burst emission.
Koshut \etal\ (1998) reported on the energy and intensity dependence
of the phase lag, or the lack thereof.

In spite of the stimulating results on the timing analysis
summarized above, only limited spectral information has been 
reported so far (Swank \etal\ 1996).
We therefore performed ASCA observations with a better spectral 
resolution than so far reported.
This paper presents the results putting particular emphasis on the
spectral features, and discusses the nature on this enigmatic X-ray object.
Preliminary analysis of a part of the ASCA observations has been
reported by Dotani \etal\ (1998).

\section{Observations and Data Reductions}

The ASCA satellite carries two Gas Imaging Spectrometers 
(GIS2 and GIS3) and two 
Solid state Imaging Spectrometers (SIS0 and SIS1) at the foci of the 
nested thin foil mirrors (Tanaka \etal\  1994).
The GIS covers a wider field of view (50' in diameter) 
with high time resolution in 
the energy range of 0.7--10 keV (Ohashi \etal\  1996, Makishima \etal\  1996),
while the SIS achieves high energy resolution (2~\% at 6~keV), 
but the field of view is smaller (11' $\times$ 11' in 1CCD mode, 
Burke \etal\  1994).
These two types of detectors have complementary characteristics.

ASCA observations of \gro\  were made twice so far.
The first  was a TOO
(target of opportunity observation) made on 1996 Feb.\ 26--27 
(hereafter observation I), and the second observation was made 
on 1997 Mar.\ 16 (observation II)\@.
Serendipitously, \gro\  was in the ASCA field 
of view while in outburst (observation II), as the primary
target of the observation was the Galactic center region. 
In addition, a sky region
including \gro\ was observed in September 1995, 
two months before the onset of the first outburst.
However, the source was not detected;
a $3\sigma$ upper limit of the source flux is listed in Table~1.

In observation I, SIS was operated in 1-CCD 
bright mode and GIS in PH mode.
The bit assignment of GIS telemetry was changed to achieve higher 
time resolution (2~msec in high bit rate) at the sacrifice of the rise 
time information of the detector signals.
Operation made of SIS in observation II was 4-CCD 
faint mode, and that of GIS was PH mode with the same 
bit  assignment as observation I.
In observation II,   \gro\  was not the primary target,
hence  was located at the  edge of the SIS fields of view.
This makes it very difficult to analyze the SIS data of \gro.   Furthermore, the SIS performance exhibited  large degradation in 4-CCD 
mode due to the radiation damage. Therefore, we analyzed
only the GIS data in observation II.

We  screened the raw data using the following criteria.
The criteria reject (1) data obtained at low elevation angles
($<\!5^\circ$) from the earth limb, (2) those affected by the 
South Atlantic Anomaly, (3) SIS data obtained at low elevation
angle ($<\!20^\circ$) from the day earth, and (4) data obtained  
at medium and low telemetry bit rate.  The 4th criterion, which
is usually not applied, was necessary because \gro\ 
was very bright during the observations.
The count rate of the source largely exceeded the telemetry
capacity in medium and low bit rate.
Hence, we used only the data in high bit rate, where the telemetry
capacities of 128 cnts/sec/GIS and 256 cnts/sec/SIS  were  large
enough to transfer all the event data of the persistent emission.
Although the data thus screened were enough for most of the 
subsequent analysis, SIS data during the bursts could not be
used in observation I, because the large increase of the count rate 
caused telemetry saturation. Thus SIS data were used only for the analysis of the 
persistent and dip emission in observation I\@.
The journal of the observations is given in Table 1.

The source photons (event data)  were  extracted from  a $3^\prime$ radius region  centered on the position of \gro\  for each detector.
Background data are accumulated from an annulus of $3^\prime-5^\prime$ radius centered 
on the source.   Figure 1 shows the GIS images of \gro\ in observations I and II, together with the regions of source and background extraction.

Since \gro\  was very bright during the ASCA observations, SIS was heavily suffered by photon pile-up (more than 2 photons fell in a pixel within a readout cycle)  near the image
center, where the flux density is large.  Thus the SIS source data  in the inner region of a radius of 1.5~arcmin are  excluded from the extraction radius of $3^\prime$.   To these data, we further corrected for  the photon pile-up using the method described in Ebisawa \etal\ (1996), although the correction factor was found to be less than 1 \%.

The large flux of \gro\ also caused significant  dead time on the on-board processing of GIS source data (event data).    Using the method described in Makishima \etal\ (1996),  the dead time fractions are estimated to be  $\sim\! 30$~\% in observation I and $\sim\! 2$~\%  in observation II in high bit rate.

The monitor data  provide further information in addition to the event data.   The most important monitor data used in this paper are  L1 data (Ohashi \etal\  1996); it is the total number of X-ray event data
(0.7--10 keV) after on-board screening.
The L1 data  are essentially  free from the telemetry limit
and suffer  from only very little dead time (25 $\mu$s).  Although the L1  data have no spatial resolution,  this is not a problem for the \gro\ case, because the source  was  a single dominant X-ray source in the  GIS field of view (see  figure~1).   Accordingly, the L1 data provide  accurate flux of GRO J1744+28,
hence can be used for the correction of  the  dead time on the event data.

The minimum time resolution for the event data and L1 data are
2~msec and 125~msec, respectively, in high bit rate mode 
in both of the observations.
Further details of ASCA and its instrumentation can be found in
Tanaka \etal\ (1994), Burke \etal\ (1991), Serlemitsos \etal\ (1995),
Ohashi \etal\ (1996) and Makishima \etal\ (1996). 

\section{Analysis and Results}

\subsection{Burst and Dip profiles}

We first  made  the X-ray light curve using the L1 data of GIS and 
they are given in Figure~2.
After the correction for the vignetting, the persistent
flux was found to be  on average $\sim\! 200$ counts/sec 
and $\sim$ 60 counts/sec for observations I and II, respectively.

We detected 12 bursts, 10 giant and 2 small bursts, during $1.6\times10^4$~sec 
exposure in observation I, and 17 giant bursts during $3.8\times10^4$~sec 
exposures in observation II\@.
Thus the mean burst intervals (for giant bursts) are 27~min for 
observation I and 37~min for observation II\@.
Although the flux level of the persistent emission differed by a 
factor of 3.3 between
the two observations, the burst rate decreased only by a factor of $\sim1.4$.
The so-called $\alpha$ value, ratio of the average luminosity
emitted in the persistent emission to that emitted in a burst
(average is taken over the time interval since the previous 
 burst) is roughly the same, $\sim\! 100$ for both observations.

In the light curve of observation I, we also found  small bursts with a peak flux 10~\% of the giant bursts.
Typical profiles of the giant and small bursts are plotted in Figure~3.
We see clear intensity dip after both types of the bursts, 
whose duration depends on the burst fluence.
We estimated the burst and dip total counts as follows.
Because the persistent flux level just before the burst was found to be
constant, we could define the pre-burst flux level for each burst.
Then, the total burst/dip counts are calculated as the sum of 
deviation of the count rate from the pre-burst level.
Not all of the burst-dip pairs were fully observed until the flux level of
the dip recovered to the persistent level, hence 
we were able to estimate the total counts of bursts and dips for
7 (observation I)  and 8 (observation II) burst-dip pairs.  
The correlation between the total counts of burst and dip is
plotted in Figure~4 for each burst-dip pair.
We see that the ratios of the burst fluence to the flux deficiency 
in the following dip are about 1/2 and 1/4 for observations I and II,  respectively.

\subsection{Pulse Profiles}

Using the L1 light curve, pulse periods in the persistent phase of each  observation were determined.
The barycentric pulse periods (corrected for the binary motion) in the
persistent phase are found to be 0.467059(2) sec and 0.4670469(8) sec, 
respectively, for observations I and II\@.   The value in the parenthesis
represents error in the last digit of the pulse period.

The pulse profile in  the persistent phases is sinusoidal, with
systematically  larger pulse fraction in observation I than that of observation II: $7.8\pm0.1$~\% and $5.7\pm0.2$~\% in the 
0.7--10~keV band for observations I and II, respectively.

During the dip and burst phases, the pulse profiles have essentially
the same shape (sinusoidal) as those of the persistent phase.
Pulse amplitude sometimes become very large (a few tens \%)
during the bursts, although the average amplitude is only slightly 
larger than that of the persistent phase.
An example of the large pulse amplitude during bursts is shown in Figure~5.

Next, we investigated the energy dependence of the pulse profile using the  X-ray event data in observation I.
 Although the event data  suffered from a significant dead time, we did not apply the dead time correction to the raw event data,  because the time resolution of the L1 data, which is necessary for the dead time correction,  is not good enough (125 msec) compared with the pulse period of 467 msec.   The pulse profiles are given in figure 6 (upper panels)  for three energy bands.  The pulse profile  are almost sinusoidal at every energy bands.  

The pulse profile of observation I is fitted with a sinusoidal function plus a constant, and  pulse fractions  defined as the ratio of the pulse  amplitude to the non-pulsed  component are calculated and are given in figure 7 with cross symbols.   To the pulse fractions,  we estimated the dead time effect using the method given in Appendix.   
The solid line in figure 7 is  the dead time corrected 
pulse fraction.   
We see  that the large difference of the pulse fraction between
higher and lower energy bands in the raw event data (crosses)
is mainly due to the dead time effect and  that  
the difference of the pulse fraction is much reduced 
after the dead time correction (solid line).

The pulse  fraction in observation II, where the dead time is negligibly small,   shows  the same dependence on the energy band as observation I,  as is  seen in the lower panel of figure 6.

\subsection{Spectral Analysis}

The spectral analysis  of \gro\ are performed  separately
for the persistent, dip and burst emissions.
As explained in section 2, both sets of SIS and GIS data are used for 
observation I (except for 
the bursts, for which only the GIS data are used), but only GIS data
are used for observation II\@.
	For each   GIS spectrum  dead time  was corrected according to the method given in Makishima \etal\ (1996).  We also corrected for the GIS gain shift, because  the GIS gain is known to increases slightly under the high  count rate (Makishima \etal\ 1996).  The gain correction factors we adopted for the burst and persistent/dip 
phases, were 0.6~\% and 0.2~\% for observation I and II, respectively.     Note that the persistent emission is not subtracted from the burst energy spectra.

We found that the energy  spectra between observations I and II
are very similar for the persistent, burst, and dip phases.
Therefore, we first try to reproduce the energy spectra of the
persistent emission, which have the best statistics, and then
applied the same model to the energy spectra of the
burst and dip phases.
Among the simple models (such as a power law, a thermal Bremsstrahlung,
and a blackbody), a power law ($\Gamma \sim 1.0$) modified by the cold 
matter absorption could reproduce the overall shape of the energy spectra
relatively well.
However, the model is rejected because of the large residual structure
around 6--7 keV as shown in the middle panel of Figure 8.
As a working hypothesis, we assume that it corresponds to
an iron emission line and we added a model of a gaussian line 
to the power law continuum.
As shown in the lower panel of Figure 8, the residual structure at 6--7 keV is
greatly reduced.

From  Table 2,  we see  that model parameters obtained 
from different sensors are not consistent with each other;  the parameters of  all the 4 sensors show 
systematic difference which is larger than the statistical errors.
Differences of the best-fit parameters are significant not only
between GIS and SIS, but also among the same sensors.
This may be partly due to a problem of the gain uncertainty of the sensor,
because the discrepancy between the data and the model is especially
large at the energies where the detection efficiency changes
rapidly (eg.\ energies at gold M-edge, silicon K-edge, xenon L-edge).
However, the discrepancy
at these energies can not be removed by simply  adjusting  the gain.
Therefore the systematic errors  is attributed not only to 
 a possible gain uncertainty, but  also to other calibration errors of 
the sensors.
Because \gro\ is a highly absorbed source having a significant
flux only in the higher energy part of the ASCA band,
calibration errors of the sensors could be larger than usual.
Since  we have no data to estimate the calibration errors
quantatively, we  regard that the parameter
differences among the sensors are the practical range of  the systematic errors.

We also fit the energy spectra of the burst and the dip phases with
the same model; results are summarized in Table 2.
Because the values of photon index and $N_{\rm H}$ couple together,
we investigate the confidence contour between these two values and find that
there is a tendency that energy spectrum becomes slightly hard 
during the bursts and slightly soft during the dips in 
observation I\@.
Although such tendency is not clear in observation II, it may
be due to poorer statistics of the data.
If we compare observations I and II, the photon index
of observation I seems to be systematically larger than
that of observation II\@.

\subsection{Iron feature}

	We found that the 6-7 keV structure is relatively well reproduced by a phenomenological model of a broad  gaussian line with the center energy of at iron  K-shell transition;
the structure is consistent with a broad line ($\sigma \sim (6-8)\times10^2$ eV) centered at  6.6--6.8 keV\@.
   In this subsection, we further apply a more physical model. 
The large line width can not be due to the thermal broadening,
because the corresponding temperature would exceed 1 MeV, but
most probably reflects the bulk motion of the line emitting gas.
If the mass accretion occurs via an accretion disk, the
line width may arise from the Doppler broadening due to the
Kepler motion of the accreting matter.
To check this possibility, we try to fit the iron feature
using the so-called disk line model (the model ``diskline'' in XSPEC)\@.
In the fit, we fixed the outer disk radius, which is hardly
constrained by the data, to 1000 $R_g$ ($R_g$: gravitational radius, GM/c$^2$).
The results are summarized in Table 3.

The disk line model is found to reproduce the iron structure well.
The ratio of the observed energy spectra to the best-fit model 
function is plotted in Figure 9.
The line energy ranged over 6.2--6.6 keV in observation I.
Therefore, we conclude
that the iron should be in a low ionization state, although the large
systematic errors do not exclude a possibility of  He-like iron.
The inner radius of the accretion disk locates around 
$10-30 \; R_g$, i.e.\ $(4-12) \times 10^6$ cm, and the
inclination angle of the accretion disk is $>\!40$ degree.
In observation II, we could not constrain the disk inclination.
If we fixed it to $15^\circ$,
a high line energy, 6.7--7.0 keV, seems to be preferred.
However, if the inclination is larger, lower line energy is still
accepted.

With the preliminary analysis,  Dotani \etal\ (1998) reported that
the  broad emission line structure was  reproduced by  a partial covering model.  Thus we  investigate  this model using the more elaborated spectra.   The results of the fitting  are  summarized in Table 3 and the ratio  to the best-fit model is shown in Figure~10.
From the shape of the residual structures in Figure~10, 
the profile of the partial covering model seems not
to match the observed shape of the iron structure very well.
Residual structures are prominent around 7~keV\@.
The fits are also generally not very good( red-$\chi^2$ of 2--3 is
obtained in observation I)\@.
Thus we conclude that the partial covering model is not preferred 
over the disk line model.

\section{Discussion}

\subsection{Interstellar Absorption and Source Distance}
 
We found that N$_{\rm H}$ values are constant at $(5-6)\times10^{22}$ 
H~cm$^{-2}$
regardless of the large luminosity changes between observations I and II.
This value is slightly larger than the one reported by Giles \etal ~(1996)
with the RXTE PCA data; their analysis, however, was preliminary.  
The stable N$_{\rm H}$ implies that the column density can be attributed to 
the interstellar absorption.
Sakano \etal\ (1997) obtained  N$_{\rm H}$ values from many X-ray sources
near the Galactic center direction and found that most of them show 
systematic dependence on the Galactic latitude, which supports that 
they really lie near the Galactic center.
The N$_{\rm H}$ value of \gro\  determined by the present work, lies
at the edge of the above column density distribution, hence we conclude that
\gro\  is really located near the Galactic center at 8.5~kpc distance.

\subsection{The Pulsed Emission}

X-ray pulsations of \gro\ are found to be very different from that of typical binary pulsars.   
The pulse amplitude is small, 6--8 \% in 2--10 keV range, and
has almost perfect sinusoidal profile.
The pulse profile of binary pulsars is usually energy dependent and 
sometimes has complex features in a lower energy band.
Such energy dependence is not recognized in \gro,
except for the increase of pulse fraction toward higher energies.
The differences in the characteristics of the X-ray pulsation
of \gro\ may result from the different parameters of the system, e.g.\ 
mass accretion rate, accretion geometry, parameters of the 
neutron star, etc.\  from typical binary pulsars.
X-ray pulsation is produced by the beamed X-ray radiation 
from the polar cap regions of the neutron star, onto which
the accreting matter is channeled by the strong magnetic field.
The sinusoidal pulse profile of \gro\ indicates that either the  emitting regions  near the polar caps are large , or the system is low inclination, as argued by
Finger \etal\ (1996), both of which will produce only weakly beamed radiation.
Small amplitude of the pulsation is also expected from a 
weakly beamed radiation.

It is considered that the size of the polar cap emission  reflect the size of the Alfv\'{e}n radius.
Accreting matter, either in a disk or in wind, is frozen in the
magnetic field at the Alfv\'{e}n radius and drifts inward along the
magnetic fields lines.
 If we assume dipole magnetic field, smaller Alfv\'{e}n radius 
makes a larger  polar cap emission.
This means that the surface magnetic field of \gro\ may be
smaller than that of typical binary pulsars.
The large mass accretion rate of \gro\ also makes the
Alfv\'{e}n radius smaller.
Several estimates of the magnetic field of \gro\ in fact
prefer a smaller value.
Finger \etal\ (1996) estimated the magnetic field to be
$<\!6\times10^{11}$ G from the assumption that the Alfv\'{e}n radius
is smaller than the co-rotation radius when the pulsar spins up and
shows pulsation.
Bildsten and Brown (1997) obtained a tighter constraint of
$<\!3\times10^{11}$ G from the same assumption, using a 
different set of data.
Cui (1997) observed evidence of the propeller effect 
when the source flux decreased and estimated the magnetic field 
to be $2.4 \times 10^{11}$ G\@.
Stark \etal\ (1998) used the same data and obtained $1.5\times10^{11}$ G\@,
assuming the presence of a contaminating source nearly at the position of GRO J1744-28. 
All these estimates favor a weak magnetic field.

We found that the pulse fraction increases with energy, which indicates that the energy spectrum of the pulsed component is harder that of the non-pulsed component, and that
the pulsed emission comes from a different region from that  of the non-pulsed component.  Since the pulsed emission is most likely from   the  polar caps of the neutron star,  the non-pulsed emission may be attributed to a larger emission region, e.g.\ whole neutron star surface.
Presence of the two emission region indicates
the presence of two accretion paths as indicated in Cui \etal\ (1997).
Some part of the accreting matter may not follow the magnetic
field and accrete spherically on to the neutron star.
This may be related to the weak magnetic field.
Presence of two accreting paths is also suspected 
in the other type II burster, the Rapid Burster, 
based on the observations of quasi-periodic oscillations
(Dotani \etal\ 1990).
Two accretion paths may, therefore, be common to the neutron star
which produces type II bursts.

We found that the pulse fraction during the bursts can be 
very large compared to that in the persistent emission.
The light curve of individual burst indicates a pulse 
fraction occasionally as large as 50 \%.
On the other hand, the pulse fraction during the dips is almost
the same as that in the persistent emission.
The small pulse fraction in the persistent emission means that
most of the matter accretes spherically on to the
neutron star and only a small fraction of matter is channeled
to the polar caps.
The large pulse amplitude during the bursts  means that some excess mass  tends to accrete  to the polar 
caps and may indicate the location of the reservoir responsible
for the bursts.

\subsection{Bursts and Dips}

We find that the burst fluence and the integrated flux deficiency
in the subsequent dip show a good correlation.
The burst fluence was approximately 1/2(observation I)  -- 1/4(observation II) of the flux deficiency.  We interpret this correlation that the burst luminosity is compensated
by the following dip luminosity, and the long-term energy release 
rate, or accretion rate is in fact nearly constant.

Therefore the  burst activity may not be due to a global increase of the
mass supply from the companion, but due to a sudden increase of the 
infall matter which has been accumulated in a reservoir 
near the neutron star.
After a burst, a fraction of the accreting matter is accumulated into the
reservoir, hence creating the flux dip phenomena.

The fact that  the burst fluence and the flux deficiency differ by a
factor of 2 (observation I) --4 (observation II) suggest a possibility that the  X-ray radiation from \gro\ is  un-isotropic, 
and the un-isotropy may change  from observations I and II, as well as from the persistent to the bursts. This possibility is also discussed in the next subsection,  in the context of super Eddington luminosity.

The $\alpha$-value is found to be $\sim$100 in both observations I and II
in spite of a factor 3--4 changes in the mass accretion rate.
This constancy is retained by the proportionality of the luminosity of
individual burst to the persistent luminosity, rather than the duty ratio
(burst interval) of the burst as indicated in Jahoda \etal\ (1998). 

\subsection{X-ray Luminosity}

Soon after the discovery of \gro, it was noticed that its X-ray
luminosity can largely exceed the Eddington limit of 
a neutron star (Giles \etal\ 1996; Jahoda \etal\ 1998).
The luminosity estimate of course depend on the source distance
and might be reduced if \gro\ were much closer. 
However, we have shown that the source column density, 
$\sim\! 5.5 \times 10^{22}$  cm$^{-2}$,
is fully consistent to the location near the Galactic center when
compared to the previous ASCA results (Sakano \etal\ 1997).

The super Eddington luminosity is, therefore, also confirmed by the 
ASCA observations.
During observation I, the luminosity (2--10~keV) of the persistent emission was
$1.6-2.0)\times10^{38}$ \ergs, 
which is comparable to the Eddington limit of a neutron star. 
Because the burst peak flux was more than 10 times the
persistent flux, the X-ray emission of \gro\ during the bursts
by far exceeds the Eddington luminosity.
In the early phase of the outburst, the X-ray flux was 10 times larger
than that of observation I, hence exceeds the Eddington luminosity 
by two orders of magnitude during the bursts as already noted by many authors.

Since the super Eddington luminosity is unlikely in rather long 
duration of the burst peak, as well as the persistent flux in the  early time of the outburst,  one may argue that the apparent super Eddington luminosity by two orders of  magnitude can be explained
by  highly un-isotropic radiation from \gro.    However as we already suggested,  most of the emission is  non-pulsed component,
and  is likely come from the whole neutron star surface, hence highly un-isotropic radiation is  unlikely, although small  un-isotropy would be possible ( see section 4.3).   Thus  we argue that  the Eddington luminosity can not simply be explained by un-isotropic radiation,  hence would remain to be an important issue  for further study.

\subsection{Iron Feature}

We find a significant iron feature in the energy spectra of
\gro\ in both observations I and II\@.
Although it is known that the diffuse Galactic ridge emission
contains an iron emission line, contamination to the energy
spectra of \gro\ is considered to be negligible.
The Fe-line intensity of the ridge emission at the location of \gro\ 
is $\sim\!20$ photons/sec/cm$^2$/sr (Maeda 1998).
This produces the contamination of $5\times10^{-5}$ photons/sec/cm$^2$ 
during the observations of \gro.
Because the line flux of \gro\ was $4\times10^{-2}$ photons/sec/cm$^2$ 
and $7\times10^{-3}$ photons/sec/cm$^2$ for observations I and II, 
respectively, contamination from the ridge emission is at least two orders 
of magnitude smaller and is completely negligible.

When fitted with a gaussian line model, the structure in observation I
is consistent with a broad line ($\sigma \sim (6-8)\times10^2$ eV) centered 
at 6.6--6.8 keV\@.
The equivalent widths are approximately 200--400 eV\@.
A disk line model also represents relatively well the iron structure.

A  problem on the disk line model, however, is that 
the best-fit parameters in observation I, especially the inner disk
radius, $(4-12) \times 10^6$ cm, and the disk inclination,
$>\!40^\circ$, do not seem to fit the \gro\ system parameters.
Magnetic field strength of 2$\times$10$^{11}$ G and X-ray luminosity
of $\sim$ 10$^{38}$ erg/s indicate an Alfv\'{e}n radius of $\sim$ 5 $\times$
10$^{7}$ cm, which is an order of magnitude larger than the above estimate.
Furthermore, we find that the inclination of the accretion disk is much larger
than the previously estimated one (Finger \etal\ 1996). 
Sturner and Dermer (1996) discussed the possible nature of the
companion star and estimated the orbital inclination to be $<18^\circ$.
Rappaport and Joss (1997) conducted a series of binary evolution
simulations to constrain the orbital inclination between $7^\circ-22^\circ$.
These estimates are much smaller than the present results.
If we force larger inner disk radius and lower inclination
angle to the diskline model, the model line profile becomes
much sharper than the observed profile and does not fit the data.

Therefore we are forced to consider additional  broadening mechanism on the conventional disk line model.   Are  the iron line  photons  Compton scattered by a surrounding high temperature plasma ?
Does the accretion disk exist inner region than the  Alfven radius, which is estimated from a spherical mass accretion ?
At this moment we have not enough data to judge  what mechanism is really working, hence this issue is open for future study.

We find that the energy dependence of the pulse amplitude 
shows a local minimum at 6--7~keV\@.
This local minimum may be related to the iron structure
of the energy spectrum.
If we assume that the iron structure is due to the broad
emission line and the line shows no pulsation, a local
minimum in the pulse amplitude is expected just as 
observed in Figure~7.
Using the equivalent width of the line, 200~eV, we estimate 
the contribution of the line flux to the total flux at 6--7~keV
bin (bin width is 1~keV) to be  roughly 20~\%.
Thus the fractional pulse amplitude may decrease by 20~\% 
in 6--7 keV\@.
In fact, the observed amplitude was 6~\% in 6--7~keV, whereas
the interpolated amplitude is 8~\%.
Thus the decrease of the pulse amplitude in 6--7~keV may be
due to the presence of the iron line which is not pulsated.

\section{Summary}

We summarize the results obtained from the ASCA observations 
of \gro\ as follows:

\begin{enumerate}
\item
We detected 10 and 17 giant Type II bursts during Observation I and II, with  
mean burst intervals of about 27 min and 37 min, respectively.     
The burst fluence is found to have good correlation with the
total flux deficiency in the following dip.
This correlation is interpreted that the average mass accretion rate is 
constant, but accretion instability makes the bursts and dips.
The burst intervals do not change very much in spite of a factor of 4 
decrease in X-ray luminosity between observation I and II\@. 
Thus the burst fluence also decreases in accordance with the
persistent flux.

\item
The absorption column is found to be constant at 
$N_{\rm H} = (5-6) \times 10^{22}$ H~cm$^{-2}$ regardless of the observation date 
and the source status (persistent, burst and dip). 
This strongly indicates that the column density corresponds to the interstellar
absorption, and the source is actually located near the Galactic center,
at a distance of 8.5~kpc. 

\item
 The persistent  X-ray luminosity in the first and the second observations are  
$(1.6-2.0) \times 10^{38}$ erg s$^{-1}$  and  $(4.1-4.2) \times 
10^{37}$ erg s$^{-1}$, respectively.  The burst peak fluxes at 
the first observation exceed the Eddington limit of a neutron star  
by a factor of 10, if the radiation is spherically symmetric.  

\item
The energy spectra of \gro\ are represented approximately by an absorbed
power law with a broad line at $\sim$6.7 keV\@.
The shape of the spectra are almost the same for the different observation
dates and types (burst, dip and persistent emission).
However, the spectra is slightly harder in observation II ($\Gamma \sim 1.0$)
than in observation I ($\Gamma \sim 1.2$), and is harder during bursts 
than in persistent phase in each set of observations.
        
\item
The energy spectrum of the pulsed component is harder than
that of the non-pulsed component.
We consider that the different spectral hardness indicates 
different emission regions, e.g.\ pulsed component from the
polar caps and non-pulsed component from the whole neutron star
surface.

\item
The presence of the iron feature is clearly seen in all energy spectra and
is also indicated by the decrease of pulse fraction at 6--7 keV\@.
The feature is well reproduced by a disk line model.
However, some line broadening mechanism is needed 
to make the disk line parameters consistent to the system
parameters of \gro.

\end{enumerate}  

\vspace{2cm}

\acknowledgments

We express our thanks to all the ASCA team members for many efforts of the 
fabrication of the satellite, launching, daily operation, soft ware developments 
and calibrations.   M.N. and Y.M. are financially supported by the Japan
society for the promotion of science.

\newpage

\section{Appendix}
Dead time of ASCA GIS is energy independent, hence does not change  the 
shape of energy spectrum,  only decreases the  normalization factor.  Therefore
the dead time effects  can simply be corrected by the L1 counts.
However,  if the  flux is time variable,  the dead time effects on the time variable component become energy dependent due to a "cross-talk" between different energy bands. This appendix
investigated the effect on the pulse fraction  with different energy bands.

Makishima \etal\ (1996) reported that  GIS has  three kinds of dead time; 
dead time caused by the
hard-wired electronics, that due to the on-board CPU processing,
and that due to a limitation in the telemetry capacity. 
These dead times  depend on the GIS mode and telemetry bit rate.
In the case of PH mode with high bit rate,  the same mode as  the analysis of 
present paper, on-board CPU processing time is the dominant source of the dead time.
Therefore, in this appendix, we consider only the CPU-induced dead time; the dead time fraction ,  $f_{\rm CPU}$, can be expressed as 
$f_{\rm CPU} = 1-(y+{\rm e}^{-y})^{-1}$ with $y = \tau \cdot X$, 
where $\tau $ is the average CPU time per event
($\tau = 8.1$ msec; Makishima et al.\ 1996), and $X$
is the incident photon flux in the total energy band.
When the incident flux exceeds a few tens photons per second, 
dead time effects of GIS become significant.

We consider a simple case that the time variation (pulsation) of the incident
photon flux $x(t;E)$ is represented by a sinusoidal form:
\begin{equation}
\label{eq:append1}
x(t;E) = f_0(E) + f_1(E) \sin(\omega t),
\end{equation}
where $f_0(E)$ is a constant flux and $f_1(E)$ is the amplitude of
sinusoidal variation.
Since  any time variations of X-ray flux  can be expressed by Fourier series,  results obtained in this appendix can be applied to  general X-ray variabilities.

%
%

Since the dead time fraction  of  GIS  does not depend on  energy,
detected (dead time included) count rate $x_d(t;E)$ at energy $E$ is given   as:
\begin{equation}
\label{eq:append2}
x_d(t;E) = \frac{f_0(E) + f_1(E) \sin(\omega t)}
                {X(t)\tau + \exp(-X(t)\tau)},
\end{equation}
Then  the time-averaged  count rate is calculated as
\begin{equation}
\overline{x_d}(E) = \frac{1}{T} \int_0^T x_d(t;E) \, dt,
\end{equation}
where $T$ ($=2\pi/\omega$) is the period of sinusoidal variation.
Note that time variation of $x_d(t;E)$ is no more sinusoidal,
because the dead time fraction, which corresponds to the denominator
of eq.~(\ref{eq:append2}), is flux dependent; in other words the dead time 
fraction varies from the pulse peak to valley.

The rms fractional amplitude $r(E)$ of the variation is:
\begin{equation}
r(E) = \frac{1}{\overline{x_d}(E)} \left\{ \frac{1}{T} 
        \int_0^T (x_d(t;E) - \overline{x_d}(E))^2 \,dt \right\}^{1/2}.
\end{equation}

When there is no dead time (i.e.\ $\tau = 0$), $\overline{x_d}(E) = f_0(E)$
and $r(E) = f_1(E)/(\sqrt{2} \, f_0(E))$.

To demonstrate how $r(E)$ changes with the incident flux, we show the results
of numerical calculation for three cases.
We assume a power law energy spectrum,
$f_0(E) = A_0 E^{-\Gamma_0}$ and $f_1(E) = A_1 E^{-\Gamma_1}$,
and rms fractional variation in the total band is 0.1.
Photon index of the constant (non-pulsed) component is fixed to
$\Gamma_0 = 1$, but three different indices are  assumed for the
time variable (pulsed) component: $\Gamma_1 = 0.9$ (case 1), $\Gamma_1 = 1.0$ (case 2)
and $\Gamma_1 = 1.1$ (case 3).
For simplicity, we divide  two energy bands: 
1--5 keV and  5--10 keV \@.
The results are shown in Figure 11.   
Ratio of the rms amplitudes (i.e.\ hardness ratio of the amplitude of
the pulsed component) are also plotted in the figure.

As seen from the figure, rms fractional variations generally decrease 
with the increase of the incident flux.
When the constant  and the pulsed  components  have the
same spectral shape (center panels of Figure 11), reduction of the
amplitude is exactly same between the two energy bands.
Thus the dead time does not alter the hardness of the pulsed component.  However, when the pulsed component is harder (softer) than the
constant  emission, local minimum of fractional variation appears 
in the soft (hard) band,
and the fractional variation turns to increase for higher flux.
This is considered to be due to a  "cross-talk'  between the 
energy bands through the dead time.

When the incident flux is much larger than $1/\tau$, which corresponds to
the maximum detectable count rate, detected count rate in the two energy 
bands tend to show anti-correlation, because the total detectable count 
rate is limited by $1/\tau$.
This anti-correlation tends to cancel out the original variation of the 
energy bin.
The local minimum of $r(E)$ in the figure indicates that the original
time variation of the energy bin is almost completely canceled out 
by the anti-correlation from the other energy bin.
Note that this cancelation is not perfect, because the time variation
is not sinusoidal under the effect of dead time.
For larger incident flux, variation due to the anti-correlation overcomes
the original variation, and the total variation increases with the increase of incident flux.
In this extreme flux range,  time variations in the two energy bins are  anti-correlated.

	Conversely,  with  the assumptions that the pulse profiles are  sinusoidal and that both the constant  (non-pulsed) and pulsed components have power law energy spectra, we can calculate the (dead time corrected) pulse fractions from the detected (dead time included) pulse fractions.   Using the average flux of about 160 c/s of L1 data  and the photon  index for the constant component of about $\Gamma = 1.0$ (see section 3.3),  we found that the best-fit photon index of the pulsed component is $0.3\pm0.1$.    Thus  the energy dependent pulse fraction with no dead time  is calculated,  and is given in figure 7 with the solid line.

\newpage

\vspace{2cm}

\begin{description}

\item[Fig.~1]
Images of \gro\ in observations I (left panel) and
II(right panel) in the Galactic coordinate.
The contours are plotted with the logarithmic spacing.
In observation I, \gro\ was detected at the center of the GIS field 
of view, but in observation II near the edge of the field of view.
Because of the distortion of the point spread function of XRT near
the edge of the field of view (Serlemitsos \etal\ 1995),
the source image in observation II has elongated morphology.

\item[Fig.~2]
X-ray light curves of \gro\ during observations I (upper
panel) and II (lower panels).
The light curves are calculated using the monitor data (L1),
which cover 0.7--10 keV, and are corrected for the vignetting.
Horizontal axis indicates the ASCA time (elapsed seconds since
1993/1/1 0:00:00 (UT)).
Data gaps are due to the earth occultation of the source, satellite
passage of the South Atlantic Anomaly or the lower telemetry bit rate.

\item[Fig.~3]
An example of the X-ray light curve in 0.7--10 keV which contains both
a giant and a small burst followed by an  intensity dip.  
The light curve is GIS2
L1 data in the observation I, and is plotted with 1 sec time resolution.
Horizontal axis indicates the time since the beginning of the observation.
  
\item[Fig.~4]
Correlation between the total burst counts and the total flux deficiency 
in the subsequent dip for observations I and II\@. 
Data from different detectors and observations are indicated by  
different symbols.    

\item[Fig.~5]
Fine time profile of a giant burst in observation I. The pulse fraction during the burst is larger than that in quiescent.

\item[Fig.~6]
Folded light curves of the persistent emission in  observation I
(upper panel) and II (lower panel).  The dead time effects are not corrected.   The pulse profile is  sinusoidal with 
larger fractional amplitudes at higher energies.
The pulse fraction in observation I becomes smaller at lower energy 
because of the deadtime effects (see Appendix).

\item[Fig.~7]
Energy dependence of the pulse fraction(dead time is not corrected) for the persistent emission in observation I of GIS2.  
The local minimum of the pulse fraction at 6--7 keV is possibly related to the existence of iron. 
The solid line is the pulse fraction  after the dead time correction (see Appendix)

\item[Fig.~8]
Energy spectra of the persistent emission in observations I (left) 
and II (right) calculated from the GIS data with the best-fit 
model functions.
In the upper panels, observed energy spectra are plotted by crosses,  while the best-fit model function (an  absorbed power law with a broad 
gaussian line) is given with the solid  histograms.
The middle and lower panels show the fit residuals: middle panels used a model
of an absorbed power law and lower panels an absorbed power law plus a 
broad gaussian line (same model in the upper panels).
Note the large residual structures around 7 keV in the middle panels.

\item[Fig.~9]
Ratio of the persistent energy spectra to the
best-fit model  of  a power law plus a disk-line.    
The ratios are plotted for each sensor.

\item[Fig.~10]
Same as figure 9, but for a partial covering by a cold matter to the 
power law continuum.

\item[Fig.~11]

The rms pulse  amplitude fraction (upper panel)  in the 1-5 keV band (solid line) and the 5-10 keV band (broken line),   and the hardness ratio  of the pulsed component (lower panel) as a function of   incident flux in the 1-10 keV band.   The pulse profile is assumed to be sinusoidal with the fraction of 0.1 in the 1-10 keV band.  The spectrum for the constant (non-pulsed) component is assumed to be a power law of a photon index of 1.0, while those of the pulsed components are assumed to be 1.1 (left), 1.0 (center: solid  broken lines are overlapped)  and 0.9 (right).
 
\end{description}
\newpage

\scriptsize
\begin{deluxetable}{lcccccc}
\tablecaption{Journal of Observations}
\tablehead{
\multicolumn{1}{l}{} &\colhead{Start} &\colhead{End} &\colhead{Exposure\tablenotemark{a}}&\colhead{GIS} &\colhead{SIS} & \colhead{Flux\tablenotemark{d}}\\ 
\multicolumn{1}{l}{} &\colhead{(yy/mm/dd hh:mm)}  &\colhead{(yy/mm/dd hh:mm)} &\colhead{(sec)} & & & \colhead{($10^{-8}$ erg/sec/cm$^2$)}\\ 
}

\startdata
Pre-outburst  & 95/09/19 08:06  &95/09/21 21:20 & 5.6 $\times$ 10$^{4}$                  & PH mode & (F/B mode for H/M bit)\tablenotemark{b} & $<1\times10^{-4}$\tablenotemark{e}\\
Observation I & 96/02/26 10:08  &96/02/27 02:05 &1.6 $\times$ 10$^{4}$ & PH mode & Bright mode & 2.0\\
Observation II& 97/03/16 15:56  &97/03/18 07:41 &3.8 $\times$ 10$^{4}$ & PH mode & (Faint mode)\tablenotemark{c} & 0.50\\
\enddata
\tablecomments{
}
\tablenotetext{a}{Calculated from the GIS data.}
\tablenotetext{b}{The X-ray source was outside the fov of SIS.}
\tablenotetext{c}{The X-ray source was located at the very edge of the fov of 
                SIS, and data were not used.}
\tablenotetext{d}{Calculated for 2--10 keV.}
\tablenotetext{e}{Absorbed power law with $N_H = 5\times10^{22}$ cm$^{-2}$ and $\Gamma = 1.0$ is assumed.}
\end{deluxetable}

\scriptsize
\begin{deluxetable}{lcccccc}
\tablecaption{Best-fit Parameters for a Power-law + Gaussian Line Model}
\tablehead{
\multicolumn{1}{l}{}&\multicolumn{4}{c}{Observation I}    &\multicolumn{2}{c}{Observation II\tablenotemark{a}}\\
\multicolumn{1}{l}{}    &\multicolumn{4}{c}{-----------------------------------------------------------------} 
&\multicolumn{2}{c}{---------------------------------}\\ 
\multicolumn{1}{l}{} &\colhead{SIS0} &\colhead{SIS1}    &\colhead{GIS2}      &\colhead{GIS3}&\colhead{GIS2} 
&\colhead{GIS3}  \\ 
}

\startdata
\multicolumn{5}{l}{}\\
\multicolumn{5}{l}{\ \ \ \ \ \ -------- Persistent -------}\\
\multicolumn{5}{l}{}\\
$N_{\rm H}$ ($10^{22}$ cm$^{-2}$) & $5.8\pm0.1$ & $5.6\pm0.1$   & $5.4\pm0.1$    & $6.2\pm0.1$    & $5.2\pm0.1$ & $5.3\pm0.1$ \\
Photon index ($\Gamma$)    & $1.34\pm0.03$ & $1.22\pm0.02$ & $1.17\pm0.03$  & $1.38\pm0.03$ & $1.03\pm0.03$ & $0.93\pm0.03$ \\ 
Line center (keV)          & $6.66\pm0.05$ & $6.72\pm0.05$ & $6.75\pm0.08$  & $6.60\pm0.08$ & $6.63\pm0.07$ & $6.74\pm0.06$ \\
Line width ($\sigma$; keV) & $0.79\pm0.10$ & (0.7) & $0.59\pm0.13$  & $0.68^{+0.05}_{-0.10}$ & $0.34\pm0.09$ &  $<0.26$ \\
Equivalent width (keV)    & $0.40\pm0.06$ & $0.39\pm0.03$ & $0.19\pm0.04$  & $0.30^{+0.06}_{-0.15}$    & $0.16\pm0.04$  & $0.13\pm0.03$ \\  
Reduced-$\chi^2$ (d.o.f.)  & 1.62 (142)    & 2.27 (143) &0.87 (89)       & 1.20 (89)         & 1.67 (89)         & 1.93 (89)     \\
\multicolumn{5}{l}{}\\
\multicolumn{5}{l}{\ \ \ \ \ \ -------- Burst ------------}\\
\multicolumn{5}{l}{}\\
$N_{\rm H}$ ($10^{22}$ cm$^{-2}$) & \multicolumn{2}{c}{------\tablenotemark{a}} & $6.0^{+0.5}_{-0.1}$  & $6.5\pm0.5$  & $6.3^{+0.7}_{-0.4}$  & $5.9\pm0.5$ \\ 
Photon index ($\Gamma$)          & \multicolumn{2}{c}{------} & $1.07^{+0.26}_{-0.12}$ & $1.31\pm0.12$ & $1.06\pm0.13$  & $0.95^{+0.16}_{-0.04}$ \\ 
Line center (keV)                & \multicolumn{2}{c}{------} & $6.5^{+0.6}_{-0.4}$    & $6.6^{+0.3}_{-0.2}$ & $6.57\pm0.12$  & $6.8\pm0.2$\\
Line width ($\sigma$; keV)       & \multicolumn{2}{c}{------} & $<1.7$   &  $<0.8$   & (0.0)    & $<0.5$ \\
Equivalent width (keV)         & \multicolumn{2}{c}{------} & $0.11^{+0.50}_{-0.07}$  & $0.19^{+0.25}_{-0.11}$  & $0.14\pm0.07$  & $0.24^{+0.15}_{-0.12}$ \\ 
Reduced-$\chi^2$ (d.o.f.)        & \multicolumn{2}{c}{------} &1.24 (89)      &1.01 (89)         &1.04 (90)         &0.85 (89)     \\ 
\multicolumn{5}{l}{}\\         
\multicolumn{5}{l}{\ \ \ \ \ \ -------- Dip --------------}\\
\multicolumn{5}{l}{}\\
$N_{\rm H}$ ($10^{22}$ cm$^{-2}$) & $5.5\pm0.2$ & $5.5\pm0.1$  & $5.8^{+0.7}_{-0.2}$   & $6.3\pm0.3$   & $5.3\pm0.2$   & $5.3\pm0.2$  \\
Photon index ($\Gamma$)     & $1.22^{+0.12}_{-0.04}$ & $1.21\pm0.04$ & $1.30\pm0.08$ & $1.52^{+0.12}_{-0.08}$ & $1.06\pm0.05$ & $1.02\pm0.06$ \\
Line center (keV)       & $6.80\pm0.10$ & $6.73\pm0.10$ & $6.78\pm0.22$ & $6.62\pm0.18$  & $6.59\pm0.11$  & $6.79\pm0.11$ \\
Line width ($\sigma$; keV)  & $0.4^{+0.7}_{-0.1}$ & (0.7) & $0.7^{+0.4}_{-0.3}$  & $0.8\pm0.3$  & $0.4\pm0.2$ & $0.44\pm0.20$ \\
Equivalent width (keV)     & $0.19^{+0.33}_{-0.05}$ & $0.37\pm0.06$ & $0.28^{+0.21}_{-0.12}$   & $0.45^{+0.21}_{-0.16}$   & $0.19^{+0.08}_{-0.06}$   & $0.27^{+0.11}_{-0.08}$  \\
Reduced-$\chi^2$ (d.o.f.)   & 1.71 (93) & 1.77 (94) & 1.09 (89)      & 1.13 (89)         & 1.15 (89)         & 1.15 (89)       \\
\enddata
\tablecomments{The errors (upper limits) quoted  are in 90 \% confidence 
limit for a single parameter.
Parameters in the parenthesis are fixed in the fitting.}
\tablenotetext{a}{No SIS data are available.}
\end{deluxetable}

\scriptsize
\begin{deluxetable}{lcccccc}
\tablecaption{Fitting Results of the Persistent Emission}
\tablehead{
\multicolumn{1}{l}{}&\multicolumn{4}{c}{Observation I}    &\multicolumn{2}{c}{Observation II\tablenotemark{a}}\\
\multicolumn{1}{l}{}    &\multicolumn{4}{c}{-----------------------------------------------------} 
&\multicolumn{2}{c}{---------------------------------}\\ 
\multicolumn{1}{l}{} &\colhead{SIS0} &\colhead{SIS1}    &\colhead{GIS2}      &\colhead{GIS3}&\colhead{GIS2} 
&\colhead{GIS3}  \\ 
}

\startdata
\multicolumn{5}{l}{}\\
\multicolumn{5}{l}{\ \ \ \ \ \ -------- Disk line model -------}\\
\multicolumn{5}{l}{}\\
$N_{\rm H}$ ($10^{22}$ cm$^{-2}$) & $5.7\pm0.1$ & $5.6\pm0,1$ & $5.0\pm0.1$ & $5.2\pm0.1$ & $5.2\pm0.1$ & $5.2\pm0.2$ \\
Photon index ($\Gamma$)    & $1.30\pm0.02$ & $1.24\pm0.03$ & $1.20\pm0.02$  & $1.28\pm0.02$ & $1.02\pm0.03$ & $0.87\pm0.04$ \\
Line center (keV)          & $6.56\pm0.07$ & $6.51\pm0.07$ & $6.45\pm0.06$  & $6.27\pm0.04$ & $6.92\pm0.07$ & $6.79\pm0.10$ \\
R$_{\rm in}$ (GM/c$^2$) & $13\pm3$ & $15\pm3$ & $22^{+9}_{-5}$ & $29\pm4$ & $11\pm5$ & $>30$\\
R$_{\rm out}$ (GM/c$^2$) & (1000) & (1000) & (1000) &  (1000) & (1000) & (1000) 
\\
Inclination (deg)          & $48\pm6$ & $>65$ & $50^{+13}_{-8}$ & $>75$ & (15)\tablenotemark{a} & (15)\tablenotemark{a} \\
Equivalent width (keV)    & $0.31\pm0.03$ & $0.46\pm0.05$ & $0.20\pm0.01$ & $0.21\pm0.01$ & $0.16\pm0.03$ & $0.11\pm0.02$ \\
Reduced-$\chi^2$ (d.o.f.)  & 1.56 (141) & 2.32 (141) & 1.82 (90) & 1.97 (90) & 1.62 (92) & 1.90 (92) \\
\multicolumn{5}{l}{}\\         
\multicolumn{5}{l}{\ \ \ \ \ \ ----- Partial covering model -----}\\
\multicolumn{5}{l}{}\\
$N_{\rm H}$ ($10^{22}$ cm$^{-2}$) & $6.1\pm0.1$ & $6.0\pm0.1$ & $5.6\pm0.1$ & $6.5\pm0.1$ & $5.8\pm0.2$ & $6.0\pm0.2$ \\
Photon index ($\Gamma$)    & $1.54\pm0.04$ & $1.49\pm0.04$ & $1.53\pm0.03$ & $1.65\pm0.05$ & $1.35\pm0.07$ &  $1.28\pm0.08$ \\
$N_{\rm H}^{\rm Partial\;Covering}$ ($10^{22}$ cm$^{-2}$) & $52\pm4$ & $59\pm4$ & $54\pm3$ & $46\pm3$ & $77\pm10$ & $90\pm15$\\
Covering fraction & $0.44\pm0.03$ & $0.52\pm0.03$ & $0.46\pm0.02$ & $0.46\pm0.03$ & $0.55\pm0.05$ & $0.62\pm0.06$ \\
Reduced-$\chi^2$ (d.o.f.)  & 2.78 (143) & 2.25 (143) & 2.19 (92) & 2.29 (92) & 1.58 (93) & 1.73 (93)  \\
\enddata
\tablecomments{The errors (upper limits) quoted are in 90 \% confidence 
limit for a single parameter.
Parameters in the parenthesis are fixed in the fitting.}
\tablenotetext{a}{The parameter was fixed because the data could not constrain it.}
\end{deluxetable}

\end{document}